# Adaptive Nonlinear Model Reduction for Fast Power System Simulation

Denis Osipov, *Student Member, IEEE*, and Kai Sun, *Senior Member, IEEE*

*Abstract*--The paper proposes a new adaptive approach to power system model reduction for fast and accurate time-domain simulation. This new approach is a compromise between linear model reduction for faster simulation and nonlinear model reduction for better accuracy. During the simulation period, the approach adaptively switches among detailed and linearly or nonlinearly reduced models based on variations of the system state: it employs unreduced models for the fault-on period, uses weighted column norms of the admittance matrix to decide which functions to be linearized in power system differential-algebraic equations for large changes of the state, and adopts a linearly reduced model for small changes of the state. Two versions of the adaptive model reduction approach are introduced. The first version uses traditional power system partitioning where the model reduction is applied to a defined large external area in a power system and the other area defined as the study area keeps full detailed models. The second version applies the adaptive model reduction to the whole system. The paper also conducts comprehensive case studies comparing simulation results using the proposed adaptively reduced models with the linearly reduced model on the Northeast Power Coordinating Council 140-bus 48-machine system.

*Index Terms*--Linear model reduction, nonlinear model reduction, power system partitioning, power system simulation.

## I. INTRODUCTION

POWER system simulation is very important for grid operations and planning at electricity utilities. It can assess dynamic security under a certain operating condition of the studied power system following a given contingency such as loss of a transmission line or generator unit. Essentially, power system simulation is to obtain a time-series trajectory of the system state for a specified simulation window by solving the initial value problem of a set of nonlinear differential-algebraic equations co-determined by the mathematical model of the whole system, the operating condition and the contingency. Nowadays, the fast growth in electricity demand and a relatively slow construction of new transmission infrastructure are pushing power transmission systems to be operated closer to their stability limits, and motivating the transition of power system simulation from the offline planning stage to the real-time operation environment.

One way to increase the speed of simulation of a complex power grid is to conduct network partitioning and then model reduction. For example, a traditional approach defines a study area, which is an important small part of the system for dynamic security assessment, considering all the details, and reduces the model of the rest of the system, i.e. the external area. In practice, an additional buffer zone is often defined in between with moderate reduction to connect the study and the external areas [1]. The methods for model reduction on the external area can be divided into two main groups: the ones that preserve the structure of a power system and the ones that use mathematical transformations from original states to nonphysical states that are subsequently reduced.

The most widely used methods from the first group are coherency-based methods [2]-[4], which were originally developed for power system model reduction and conduct the following steps: coherency identification, aggregation of coherent generators, and network reduction. After the first step, the generators that oscillate together following a disturbance are included into one group. The groups of coherent generators are then aggregated into individual equivalent generators connected with each other by equivalent branches and with the study area by a reduced system network. This creates a unique boundary between the external area and the study area and does not allow arbitrary division between areas. In addition, if the topology of the original system changes, it can affect the coherency and consequently the split between the study area and the generator grouping of the external area. This can change the boundary between the study area and the external area. Thus, the grouping of generators based on coherency has an inherent limitation on the way a system can be partitioned.

The second group of methods does not have that limitation as the states are transformed into a new state space. Thus, the system can be partitioned in any way. These methods came from the control field of engineering. In the most used methods from this category, the external area model is first linearized and then reduced using balanced truncation [5]-[7] or Krylov subspace methods [8]-[10]. The linearization of the model gives acceptably accurate results when concerned disturbances are small but once the size of the disturbance increases the linearized model cannot guarantee an accurate representation of the original part of the system. To improve the accuracy of large-disturbance simulation, nonlinear model reduction methods can be used [11]-[15]. However, as shown in [16], application of nonlinear model reduction cannot give

This work was supported by NSF grant ECCS-1610025 and the CURENT Engineering Research Center.

D. Osipov and K. Sun are with the Department of Electrical Engineering and Computer Science, University of Tennessee, Knoxville, TN 37996 USA (email: dosipov@vols.utk.edu, kaisun@utk.edu).

substantial computational time decrease as compared to the original model. In addition, some of the methods require training simulation data to create a reduced model, which cannot guarantee adequate performance during all possible disturbances. If the disturbance is very different from that with the training set, the model reduction error can substantially increase [17].

Compared to the existing work this paper proposes a new adaptive model reduction approach, which is a compromise between linear model reduction and nonlinear model reduction in terms of accuracy and speed of time-domain simulation using the reduced model. A comprehensive study is also presented to compare this adaptive model reduction approach with the linear model reduction approach. During the simulation period, the approach adaptively switches among detailed and linearly or nonlinearly reduced models based on variations of the system state: it employs unreduced models for the fault-on period, uses weighted column norms of the admittance matrix to decide which functions to be linearized in the power system model for large changes of the state, and adopts a linearly reduced model for small changes of the state.

The version of the adaptive approach described above uses traditional topological power system partitioning with the study area and the external area. This partitioning creates an additional error that can affect the performance of the model reduction. To address the partitioning error the second version of the adaptive approach is proposed where the model reduction is performed to the whole system.

Thus, the main contributions of this paper are in the following aspects: 1) linearization of only certain functions and the selection criterion of the functions to be linearized; 2) adaptive switching among detailed and linearly or nonlinearly reduced models and the switching criterion; 3) application of power system model reduction to the system without partitioning.

The remainder of this paper is organized as follows. Section II describes the model and the approach used in power system partitioning, presents the linear model reduction approach and proposes the adaptive model reduction approach including the model reduction method as a hybrid of nonlinear and linear model reduction techniques, the algorithm enabling adaptive switching among models of different levels of details and the second version of the adaptive approach that can be applied to the whole power system. In Section III the proposed approaches are tested together with the linear model reduction approach on the Northeast Power Coordinating Council (NPCC) 48-machine 140-bus system. Finally, conclusions are drawn in Section IV.

## II. PROPOSED ADAPTIVE MODEL REDUCTION

### A. Power System Partitioning

As it was mentioned above, in power system model reduction the system is divided into two areas: 1) the study area, which is the main interest of an investigator, where all details are preserved and all disturbances are originated from; 2) the external area, which can be simplified and reduced. The partitioned power system is shown in Fig. 1.

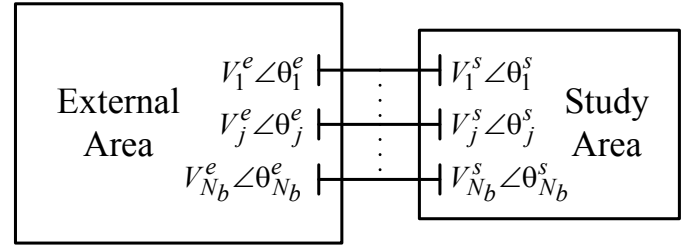

Fig. 1. Partitioned power system.

Each area of the original system is connected to other area by several tie-lines. For every area, each tie-line is treated as a simple fictitious generator with the internal voltage phasor equal to the voltage phasor of the corresponding boundary bus in the opposite area and with the armature resistance and transient reactance equal to the resistance and reactance of the corresponding tie-line. These fictitious generators are treated as constant voltage sources during each iteration and represent the electrical power injections from one area to the other area. Therefore, voltage magnitudes and voltage angles of boundary buses in one area are the inputs to the model of the other area. At every iteration of the system simulation, each area is calculated separately, then boundary bus voltages of all areas are recalculated and their values are sent as inputs to the corresponding area to perform the next iteration.

In this paper, power system generators are modeled based on the detailed two-axis machine model, the non-reheat steam turbine model, the first-order governor model and the IEEE type 1 exciter model [18]. In each area of the partitioned system every generator is described by the following nine differential equations:

$$\begin{cases} \dot{\delta}_i = \omega_{base}(\omega_i - 1) \\ T_{ch_i}\dot{P}_{m_i} = -P_{m_i} + P_{gv_i} \\ T_{gv_i}\dot{P}_{gv_i} = -P_{gv_i} + P_{ref_i} - \dfrac{(\omega_i - 1)}{R_i} \\ T_{A_i}\dot{V}_{R_i} = -V_{R_i} + K_{A_i}R_{f_i} - \dfrac{K_{A_i}K_{F_i}}{T_{F_i}}E_{fd_i} + K_{A_i}\left(V_{ref_i} - V_{t_i}\right) \\ T_{F_i}\dot{R}_{f_i} = -R_{f_i} + \dfrac{K_{F_i}}{T_{F_i}}E_{fd_i} \\ T_{E_i}\dot{E}_{fd_i} = -\left(K_{E_i} + A_{E_i}e^{B_{E_i}E_{fd_i}}\right)E_{fd_i} + V_{R_i} \\ T'_{qo_i}\dot{E}'_{d_i} = -E'_{d_i} + \left(X_{q_i} - X'_{q_i}\right)I_{q_i} \\ T'_{do_i}\dot{E}'_{q_i} = -E'_{q_i} - \left(X_{d_i} - X'_{d_i}\right)I_{d_i} + E_{fd_i} \\ 2H_i\dot{\omega}_i = P_{m_i} - E'_{d_i}I_{d_i} - E'_{q_i}I_{q_i} - D_i(\omega_i - 1) \end{cases} \quad (1)$$

where

$$I_{d_i} = \sum_{j=1}^{N_g} E'_{d_j} \left[ G_{ij} \cos(\delta_i - \delta_j) + B_{ij} \sin(\delta_i - \delta_j) \right]$$
$$+ \sum_{j=1}^{N_g} E'_{q_j} \left[ G_{ij} \sin(\delta_i - \delta_j) - B_{ij} \cos(\delta_i - \delta_j) \right]$$
$$+ \sum_{j=1}^{N_b} V_j \left[ G_{ij} \sin(\delta_i - \theta_j) - B_{ij} \cos(\delta_i - \theta_j) \right],$$

$$I_{q_i} = \sum_{j=1}^{N_g} E'_{d_j} \left[ -G_{ij} \sin(\delta_i - \delta_j) + B_{ij} \cos(\delta_i - \delta_j) \right]$$
$$+ \sum_{j=1}^{N_g} E'_{q_j} \left[ G_{ij} \cos(\delta_i - \delta_j) + B_{ij} \sin(\delta_i - \delta_j) \right]$$
$$+ \sum_{j=1}^{N_b} V_j \left[ G_{ij} \cos(\delta_i - \theta_j) + B_{ij} \sin(\delta_i - \theta_j) \right],$$

Here, $\delta_i$ and $\omega_i$ are the rotor angle and speed of generator $i$ in rad and rad/s, respectively; $\omega_{base} = 120\pi$ rad/s is the base speed; $P_m$, $P_{gv_i}$ and $P_{ref_i}$ the mechanical power, the governor output power and the reference power, respectively; $R_i$ is the speed regulation factor; $E'_{d_i}$, $E'_{q_i}$, $X_{d_i}$, $X_{q_i}$, $X'_{d_i}$, $X'_{q_i}$, $I_{d_i}$ and $I_{q_i}$ are respectively the $d$- and $q$-axis internal voltages, synchronous reactances, transient reactances and currents all in p.u.; $V_{R_i}$, $K_{A_i}$, $R_{f_i}$, $K_{F_i}$, $E_{fd_i}$, $V_{ref_i}$, $V_{t_i}$, $K_{E_i}$, $A_{E_i}$ and $B_{E_i}$ are the voltage regulator input, amplifier gain, rate feedback, feedback gain, field voltage, reference voltage, terminal bus voltage, exciter gain and exciter saturation coefficients; Time constants $H_i$, $T_{ch_i}$, $T_{gv_i}$, $T_{A_i}$, $T_{F_i}$, $T_{E_i}$, $T'_{qo_i}$ and $T'_{do_i}$ are respectively the generator inertia, turbine charging time, governor time constant, amplifier time constant, feedback time constant, the exciter time constant, the q-axis open circuit time constant and the d-axis open circuit time constant all in s; $D_i$ is the inertia and damping coefficient; $N_g$ is the number of generators; $N_b$ is the number of boundary buses; $G_{ij}$ and $B_{ij}$ are conductance and susceptance between generator $i$ and generator $j$ in p.u.; $V_j$ is the voltage magnitude at boundary bus $j$ in the opposite area in p.u.; $\theta_j$ is the voltage angle at boundary bus $j$ in the opposite area in rad.

*B. Model Reduction*

If model reduction is applied to the external area it is necessary to define states and inputs of the system. Considering that every generator is described by nine differential equations and every boundary bus has the voltage magnitude and the voltage angle as its parameters, let $n = 9N_g$ and $m = 2N_b$ respectively denote the number of states and the number of inputs of the external area and let the outputs of the system be the states of the system.

Then the system (1) can be described as the nonlinear system:

$$\begin{cases} \dot{\mathbf{x}} = \mathbf{f}(\mathbf{x}, \mathbf{u}) \\ \mathbf{y} = \mathbf{x} \end{cases} \quad (2)$$

where $\mathbf{x} = (\boldsymbol{\delta} \ \mathbf{P}_m \ \mathbf{P}_{gv} \ \mathbf{V}_R \ \mathbf{R}_f \ \mathbf{E}_{fd} \ \mathbf{E}_d \ \mathbf{E}_q \ \boldsymbol{\omega})^T$ and $\mathbf{u} = (\boldsymbol{\theta} \ \mathbf{V})^T$; $\mathbf{x} \in \mathrm{R}^n$ is the state vector; $\mathbf{u} \in \mathrm{R}^m$ is the input vector; $\mathbf{y} \in \mathrm{R}^n$ is the output vector.

*1) Linear Model Reduction:*

The system (2) can be linearized around an equilibrium point as:

$$\begin{cases} \Delta\dot{\mathbf{x}} = \mathbf{A}\Delta\mathbf{x} + \mathbf{B}\Delta\mathbf{u} \\ \Delta\mathbf{y} = \mathbf{C}\Delta\mathbf{x} \end{cases} \quad (3)$$

where $\Delta\mathbf{x}$, $\Delta\mathbf{u}$ and $\Delta\mathbf{y}$ are the deviation variables of respectively the original states, inputs and outputs; $\mathbf{A} \in \mathrm{R}^{n \times n}$ is the matrix of partial derivatives of the functions in (1) with respect to each state evaluated at the equilibrium point; $\mathbf{B} \in \mathrm{R}^{n \times m}$ is the matrix of partial derivatives of the functions in (1) with respect to each input evaluated at the equilibrium point; $\mathbf{C} \in \mathrm{R}^{n \times n}$ is the identity matrix.

The system (3) can be reduced using a linear reduction method, for example, the balanced truncation method [5]. To apply this method Lyapunov equations are solved to get controllability Gramian $\mathbf{W}_c$ and observability Gramian $\mathbf{W}_o$:

$$\begin{cases} \mathbf{A}\mathbf{W}_c + \mathbf{W}_c \mathbf{A}^T + \mathbf{B}\mathbf{B}^T = \mathbf{0} \\ \mathbf{A}^T \mathbf{W}_o + \mathbf{W}_o \mathbf{A} + \mathbf{C}^T \mathbf{C} = \mathbf{0} \end{cases} \quad (4)$$

The Gramians are then used to calculate transformation matrix $\mathbf{T}$ and its inverse $\tilde{\mathbf{T}}$.

Matrix $\mathbf{T}$ transforms the states from the original state space to a new balanced state space: $\Delta\tilde{\mathbf{x}} = \mathbf{T}\Delta\mathbf{x}$.

In the resulted new balanced system, the states are arranged in a such way that the first state is the most controllable and the most observable and the last state is the least controllable and the least observable. Henkel singular values show this relationship:

$$\sigma_1 > \sigma_2 > \cdots > \sigma_i > \cdots > \sigma_{n-1} > \sigma_n \geq 0, \quad (5)$$

where

$$\sigma_i = \sqrt{\lambda_i(\mathbf{W}_c \mathbf{W}_o)} = \Sigma_{ii}.$$

Considering the above-mentioned fact only the first $r$ states can be kept and the rest can be truncated. $H_\infty$ norm of the error of balanced truncation is bounded by the following expression:

$$\|\varepsilon\|_\infty \leq 2 \sum_{i=r+1}^{n} \sigma_i. \quad (6)$$

The transformation matrix and its inverse are recalculated as follows:
$$\mathbf{T} = \mathbf{PT}, \quad \tilde{\mathbf{T}} = \tilde{\mathbf{T}}\mathbf{P}^T, \quad (7)$$

where $\mathbf{P} = \begin{pmatrix} \mathbf{I} & \mathbf{0} \end{pmatrix}$ is the identity matrix, the last $(n-r)$ rows of which are deleted.

Thus, the balanced truncated system is represented as follows:
$$\begin{cases} \Delta\dot{\tilde{\mathbf{x}}} = \mathbf{T}\mathbf{A}\tilde{\mathbf{T}}\Delta\tilde{\mathbf{x}} + \mathbf{T}\mathbf{B}\Delta\mathbf{u} \\ \Delta\mathbf{y} = \mathbf{C}\tilde{\mathbf{T}}\Delta\tilde{\mathbf{x}} \end{cases} \quad (8)$$

The system in (8) can be written in a more compact form:
$$\begin{cases} \Delta\dot{\tilde{\mathbf{x}}} = \tilde{\mathbf{A}}\Delta\tilde{\mathbf{x}} + \tilde{\mathbf{B}}\Delta\mathbf{u} \\ \Delta\mathbf{y} = \tilde{\mathbf{C}}\Delta\tilde{\mathbf{x}} \end{cases} \quad (9)$$

where
$$\tilde{\mathbf{A}} = \mathbf{T}\mathbf{A}\tilde{\mathbf{T}}, \quad \tilde{\mathbf{A}} \in \mathrm{R}^{r \times r},$$
$$\tilde{\mathbf{B}} = \mathbf{T}\mathbf{B}, \quad \tilde{\mathbf{B}} \in \mathrm{R}^{r \times m},$$
$$\tilde{\mathbf{C}} = \mathbf{C}\tilde{\mathbf{T}}, \quad \tilde{\mathbf{C}} \in \mathrm{R}^{n \times r}.$$

*2) Proposed Hybrid Model Reduction:*

In this paper, a model reduction method is proposed as a hybrid of nonlinear and linear model reduction techniques. As shown in [19]-[20], the transformation matrices $\mathbf{T}$ and $\tilde{\mathbf{T}}$ can be used to reduce the nonlinear system as well. In this case, the system can be represented as follows:
$$\begin{cases} \dot{\tilde{\mathbf{x}}} = \mathbf{T}\mathbf{f}(\tilde{\mathbf{T}}\tilde{\mathbf{x}}, \mathbf{u}) \\ \mathbf{y} = \tilde{\mathbf{T}}\tilde{\mathbf{x}} \end{cases} \quad (10)$$

The system in (10) has fewer states than the original system but it is still necessary to compute all nonlinear functions in $\mathbf{f}$. Thus, there is basically no reduction in computation time.

To address that problem, reference [21] suggests eliminating some of the functions. However, as it is shown in [22] it can create large errors due to the model reduction.

In the proposed hybrid model reduction approach, the functions that have the least contributions to the dynamics between the external area and a study area are not eliminated but linearized. To evaluate contributions of the functions, let us consider the expressions for the d-axis current and the q-axis current of generators in the external area as these expressions have most nonlinearities and are used in 33% of all differential equations in (1).

Nonlinearities in the expressions are cosine and sine functions and coefficients of these functions are conductances and susceptances between generators including fictitious generators representing the boundary between the external area and the study area. These values are real and imaginary parts of elements of the admittance matrix:
$$Y_{ij} = G_{ij} + jB_{ij}. \quad (11)$$

The matrix can be divided into four submatrices:
$$\mathbf{Y} = \begin{pmatrix} \mathbf{Y}_{11} & \mathbf{Y}_{12} \\ \mathbf{Y}_{21} & \mathbf{Y}_{22} \end{pmatrix}, \quad (12)$$

where $\mathbf{Y}_{11} \in \mathrm{R}^{N_g \times N_g}$ is the admittance matrix representing connections between generators inside the external area; $\mathbf{Y}_{22} \in \mathrm{R}^{N_b \times N_b}$ is the admittance matrix representing connections between fictitious generators; $\mathbf{Y}_{21} = \mathbf{Y}_{12}^T \in \mathrm{R}^{N_b \times N_g}$ is the admittance matrix representing connections between the generators of the external area and the fictitious generators.

Thus, column norms of absolute values of elements in matrix $\mathbf{Y}_{21}$ can be used to determine which function to be linearized as the norms describe how close electrically each generator is to the boundary between the external area and a study area.

Column norms are calculated by:
$$\nu_i = \sqrt{\sum_{j=1}^{N_b} \left| Y_{21_{ji}} \right|^2}. \quad (13)$$

The nonlinear functions that correspond to the generators with large column norms are kept nonlinear and the nonlinear generator functions with small column norms are linearized. Thus, the hybrid reduced system can be represented as follows:
$$\begin{cases} \dot{\tilde{\mathbf{x}}} = \mathbf{T} \begin{pmatrix} \hat{\mathbf{f}}(\tilde{\mathbf{T}}\tilde{\mathbf{x}}, \mathbf{u}) \\ \mathbf{f}_l \end{pmatrix} \\ \mathbf{y} = \tilde{\mathbf{T}}\tilde{\mathbf{x}} \end{cases} \quad (14)$$

where
$$\mathbf{f}_l = \hat{\mathbf{A}}\Delta\tilde{\mathbf{x}} + \hat{\mathbf{B}}\Delta\mathbf{u} + \hat{\mathbf{x}}^0, \quad \hat{\mathbf{x}}^0 = \hat{\mathbf{P}}\mathbf{x}^0,$$
$$\hat{\mathbf{A}} = \hat{\mathbf{P}}\mathbf{A}\tilde{\mathbf{T}}, \quad \hat{\mathbf{A}} \in \mathrm{R}^{(n-q) \times r},$$
$$\hat{\mathbf{B}} = \hat{\mathbf{P}}\mathbf{B}, \quad \hat{\mathbf{B}} \in \mathrm{R}^{(n-q) \times m},$$

Vector $\hat{\mathbf{f}}$ comprises the functions that are kept nonlinear; $\mathbf{f}_l$ has the linearized functions; $\mathbf{x}^0$ is the initial state vector; $\hat{\mathbf{P}}$ is the identity matrix with deleted rows that correspond to the functions in $\hat{\mathbf{f}}$; $q$ is the number of nonlinear functions in $\hat{\mathbf{f}}$.

The system in (14) can be rewritten as:
$$\begin{cases} \dot{\tilde{\mathbf{x}}} = \mathbf{T} \begin{pmatrix} \hat{\mathbf{f}}(\tilde{\mathbf{T}}\tilde{\mathbf{x}}, \mathbf{u}) \\ \hat{\mathbf{A}}\Delta\tilde{\mathbf{x}} + \hat{\mathbf{B}}\Delta\mathbf{u} + \hat{\mathbf{x}}^0 \end{pmatrix} \\ \mathbf{y} = \tilde{\mathbf{T}}\tilde{\mathbf{x}} \end{cases} \quad (15)$$

*C. Adaptive Switching Algorithm*

Considering that the linearly reduced system gives satisfactory performance during small disturbances, the duration of a large disturbance is short, and majority of the time a system is under small or no disturbance it is reasonable to change the type of the model reduction of the external system to increase the accuracy and speed of the system simulation. Proposed adaptive algorithm changes the complexity of model reduction of the external area based on the current condition. The adaptive algorithm is shown in Fig. 2.

During the fault-on period, the original, fully detailed system model is used as the maximum accuracy of the system model is required and the duration of a fault is limited to tens of milliseconds, which does not increase much the simulation time. In the post-fault period, when the angle deviation $\Delta\delta_i$ of any generator in a study area exceeds a preset threshold $\Delta\delta_{\max}$, the external area is reduced using the hybrid model reduction method, which keeps the balance between accuracy and speed of simulation when the disturbance is large. In pre-fault and post-fault periods when all angle deviations are within the threshold, the external area is reduced using a linear model reduction method. This guarantees that most of the time when there is no disturbance or variation of the state is very small, the fastest model reduction method is applied.

To calculate the rotor angle deviation the generator with the smallest column norm is used as a reference; i.e., the reference generator is the electrically furthest generator from a study area and reacts to disturbances in study areas the least. If the time for a rotor angle deviation $\Delta\delta_i$ exceeding threshold $\Delta\delta_{\max}$, denoted by $t_{th}$, is longer than a preset limit $t_{th\max}$ a new operating condition is obtained and matrices $\tilde{\mathbf{A}}$ and $\tilde{\mathbf{B}}$ of the linearly reduced system are recalculated. This action corrects the adaptive algorithm after a large change of the system state.

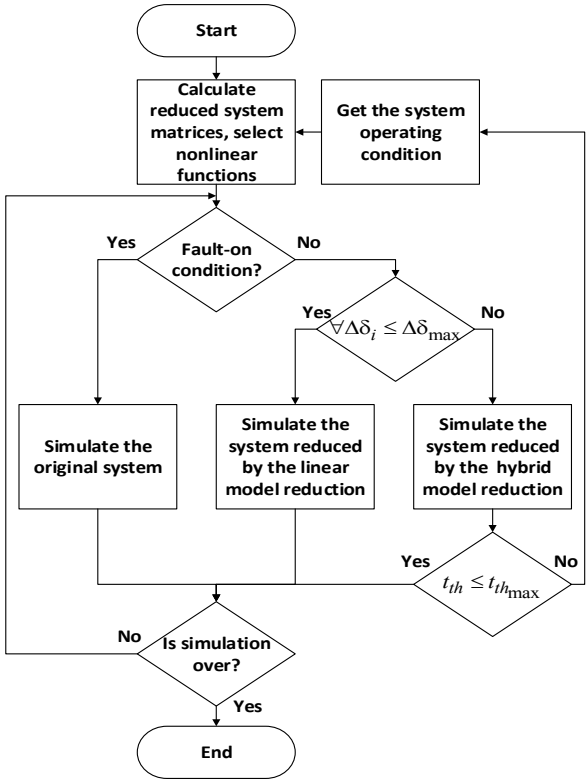

Fig. 2. Adaptive switching algorithm.

### D. Adaptive Model Reduction of the System without Partitioning

Partitioning the system into the study area and the external area creates a specific error. The error is caused by the fact that the inputs (boundary bus voltage magnitudes and boundary bus voltage angles) are calculated at the previous iteration of the simulation, i.e. the inputs are lagging by one iteration. To eliminate the partitioning error the second version of the adaptive approach is proposed. This version is applied to the whole system that is treated as just one area.

Without the partitioning, there is no need in the concept of fictitious generators representing boundary buses of the study area and the external area and the expressions in (1) for the d-axis current and the q-axis current are simplified as:

$$\begin{aligned} I_{d_i} &= \sum_{j=1}^{N_g} E'_{d_j} \left[ G_{ij} \cos(\delta_i - \delta_j) + B_{ij} \sin(\delta_i - \delta_j) \right] \\ &+ \sum_{j=1}^{N_g} E'_{q_j} \left[ G_{ij} \sin(\delta_i - \delta_j) - B_{ij} \cos(\delta_i - \delta_j) \right], \\ I_{q_i} &= \sum_{j=1}^{N_g} E'_{d_j} \left[ -G_{ij} \sin(\delta_i - \delta_j) + B_{ij} \cos(\delta_i - \delta_j) \right] \\ &+ \sum_{j=1}^{N_g} E'_{q_j} \left[ G_{ij} \cos(\delta_i - \delta_j) + B_{ij} \sin(\delta_i - \delta_j) \right]. \end{aligned}$$

As the only area of the system contains all generators including the generators from the study area whose dynamics are of the main interests, the transformation and truncation of the states are not performed and the performance improvement comes only from the linearization of nonlinear functions. In the absence of inputs from the boundary between the study area and the external area, the control matrix $\mathbf{B}$ is eliminated from (15) and the system used in the second version of the adaptive approach is simplified as:

$$\begin{cases} \dot{\mathbf{x}} = \begin{pmatrix} \hat{\mathbf{f}}(\mathbf{x}, \mathbf{u}) \\ \hat{\mathbf{A}}\Delta\mathbf{x} + \hat{\mathbf{x}}^0 \end{pmatrix} \\ \mathbf{y} = \mathbf{x} \end{cases} \quad (16)$$

where

$$\hat{\mathbf{A}} = \hat{P}A, \quad \hat{\mathbf{A}} \in R^{(n-q) \times n}.$$

All nonlinear functions representing generators of the study area are contained in $\hat{\mathbf{f}}$. The list of linearized functions corresponding to generators of the external area is the same as in the adaptive approach applied to the partitioned system described above. The adaptive switching is performed between the system in (16) and the simplified version of the linearized system in (3):

$$\begin{cases} \Delta\dot{\mathbf{x}} = \mathbf{A}\Delta\mathbf{x} \\ \Delta\mathbf{y} = \mathbf{C}\Delta\mathbf{x} \end{cases} \quad (17)$$

As all generators of the whole system including the generators of the study area are linearized in (17) the angle deviation threshold $\Delta\delta_{\max}$ of the adaptive switching algorithm is set to small value to enable switching after the oscillations are damped enough.

## III. CASE STUDIES

Comprehensive case studies are conducted to compare the proposed adaptive model reduction approaches with the traditional linear model reduction approach. A realistic power system model is tested. For the system, the study area is defined and retained with original, detailed models and the rest of the system is defined as the external area to be reduced respectively by different approaches. Then, time-domain contingency simulation using each reduced system model is conducted and compared with the simulation using the original system model. In addition, the approach is validated during different post-fault operating conditions.

*A. Temporary bus fault tests*

The linear model reduction approach and the adaptive approaches described above are applied to the NPCC 140-bus 48-generator power system. The study area is set to be the New England region of the NPCC system, which has 9 generators. The external area is set to be the rest of the system, which has 39 generators. The external area has 39×9=351 state variables and 39×4=156 nonlinear functions as the first 5 states of each generator is described by linear differential equations. The external area is connected to the study area by two tie-lines. The partitioned NPCC system is shown in Fig. 3.

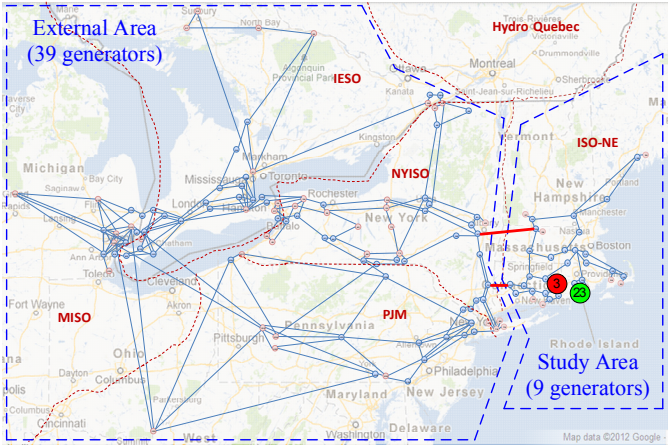

Fig. 3. Partitioned NPCC system.

The $H_\infty$ norm of the balanced truncation error is set to be equal to $10^{-5}$:

$$10^{-5} \leq 2 \sum_{i=r+1}^{n} \sigma_i. \quad (18)$$

This corresponds to truncation of 200 out of 351 state variables. To find the threshold value of the column norm of the admittance matrix that decides if generator is electrically close to the boundary a case study is performed. The system in (15) is used in this case study. The threshold is decreased with 0.1-p.u. increments from 10 until all rotor angle errors for all generators following any of the contingencies are below 6 degrees. After the case study, the threshold is set to 1 p.u. The nonlinear functions corresponding to generators with column norms of admittance matrix less than 1.0 p.u. are linearized. This corresponds to linearization of 136 out of 156 nonlinear functions of the external area in the case of partitioned system and 136 out of 192 nonlinear functions of the unpartitioned system.

The column norm threshold per unit value in not useful if the proposed approach is applied to a different system with a different base power. To make it more practical it is converted to value in siemens. The NPCC system has the base power of 100 MVA. Setting the base voltage to the common generator terminal voltage of 20 kV the new thresholds is calculated:

$$Y_{th} = 1 \frac{S_{base}}{V_{base}^2} = 1 \frac{100}{20^2} = 0.25 S \quad (19)$$

When the proposed approach is applied to a different system, the threshold is converted to a per unit value using the new system base power value and the base voltage of 20 kV.

Another case study is performed to select the angle deviation threshold $\Delta\delta_{max}$ for the adaptive switching algorithm applied to the partitioned system. The threshold is decreased with 1-degree increments from 180 degrees until all rotor angle errors for all generators following any of the contingencies are below 6 degrees. After this case study, $\Delta\delta_{max}$ is set to 67 degrees. The threshold for the adaptive approach applied to the unpartitioned system is set to 6 degrees to ensure rotor angle errors of generators representing study area are also below 6 degrees.

The simulations are performed in MATLAB R2015a on the computer with 4-GHz AMD FX-8350 processor and 8 GB of memory. The duration of simulation is set to 16 seconds, integration time step is set to 0.01 seconds.

TABLE I
COMPARISON OF ROOT MEAN SQUARE ERROR OF ROTOR ANGLE

| Bus | Gen. | CCT | Partitioned | | Unpartitioned |
|---|---|---|---|---|---|
| | | | Linear | Adapt. | Adapt. |
| 1 | 26 | 0.13 | 8.85 | 1.33 | 1.06 |
| 2 | 26 | 0.13 | 3.10 | 0.86 | 0.57 |
| 3 | 23 | 0.39 | 25.94 | 5.51 | 5.12 |
| 4 | 26 | 0.12 | 1.50 | 0.75 | 0.41 |
| 5 | 23 | 0.09 | 0.59 | 0.48 | 0.29 |
| 6 | 23 | 0.08 | 0.72 | 0.60 | 0.30 |
| 7 | 26 | 0.05 | 0.28 | 0.28 | 0.47 |
| 8 | 23 | 0.05 | 4.87 | 4.41 | 0.29 |
| 9 | 26 | 0.06 | 3.26 | 2.83 | 0.29 |
| 10 | 26 | 0.13 | 10.49 | 2.90 | 1.26 |
| 11 | 23 | 0.19 | 8.81 | 1.40 | 1.46 |
| 12 | 23 | 0.1 | 1.89 | 0.98 | 0.44 |
| 13 | 23 | 0.12 | 2.57 | 0.97 | 0.49 |
| 14 | 23 | 0.13 | 2.83 | 1.00 | 0.51 |
| 15 | 23 | 0.07 | 0.38 | 0.38 | 0.33 |
| 16 | 23 | 0.05 | 0.87 | 0.78 | 0.31 |
| 17 | 26 | 0.03 | 0.30 | 0.30 | 0.59 |
| 18 | 23 | 0.04 | 0.33 | 0.33 | 0.66 |
| 19 | 26 | 0.03 | 0.11 | 0.11 | 0.95 |
| 20 | 26 | 0.03 | 0.27 | 0.27 | 1.05 |
| 21 | 23 | 0.22 | 12.97 | 1.65 | 1.89 |
| 22 | 23 | 0.19 | 6.55 | 1.32 | 1.06 |
| 23 | 23 | 0.20 | 3.83 | 1.07 | 0.71 |
| 24 | 23 | 0.18 | 5.66 | 1.46 | 0.91 |
| 25 | 23 | 0.22 | 20.45 | 2.22 | 3.87 |
| 26 | 26 | 0.03 | 0.35 | 0.34 | 0.86 |
| 27 | 23 | 0.10 | 0.29 | 0.27 | 0.46 |
| 28 | 26 | 0.12 | 0.28 | 0.30 | 0.42 |
| 29 | 23 | 0.06 | 3.15 | 3.48 | 0.43 |
| 30 | 23 | 0.06 | 1.21 | 1.35 | 0.38 |
| 31 | 23 | 0.08 | 0.40 | 0.43 | 0.25 |
| 32 | 26 | 0.11 | 1.25 | 0.79 | 0.42 |
| 33 | 26 | 0.11 | 1.05 | 0.65 | 0.39 |
| 34 | 23 | 0.15 | 3.79 | 0.89 | 0.63 |
| 35 | 26 | 0.14 | 3.14 | 0.82 | 0.52 |
| 36 | 23 | 0.21 | 4.03 | 0.92 | 0.61 |

To compare the approach performance during a large disturbance, a three-phase short circuit fault lasting for the critical clearing time (CCT) is created separately at every bus of the study area. The errors of all outputs (state variables) of all generators in the study area are analyzed and the rotor angle state variable has the largest error for each of the generators. For every fault, the generator with the largest error of the rotor angle is found and used to compare the approaches. The results of comparison of rotor angle root mean square errors of the linear model reduction approach and the adaptive approach are shown in Table I. As it can be seen the linear model reduction approach cannot guarantee satisfactory performance during large disturbances generating the errors of more than 25 degrees while the proposed adaptive approach keeps the error for all disturbances within 6 degrees both for the case with the system partitioned into the study and the external area and the case with only one area representing the whole system.

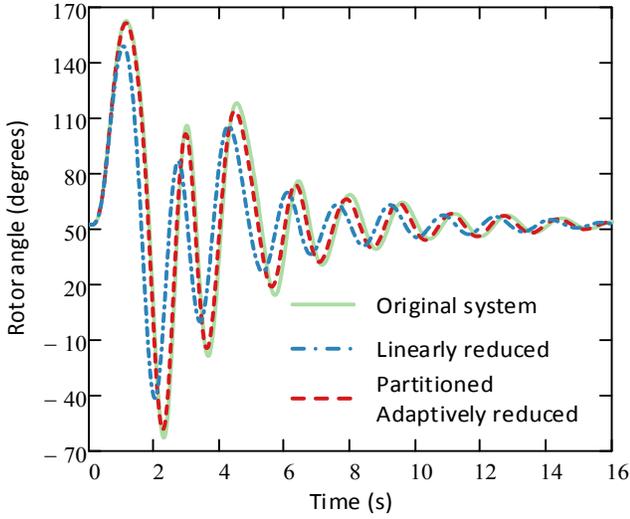

Fig. 4. Rotor angle of generator 23 following the fault at bus 3.

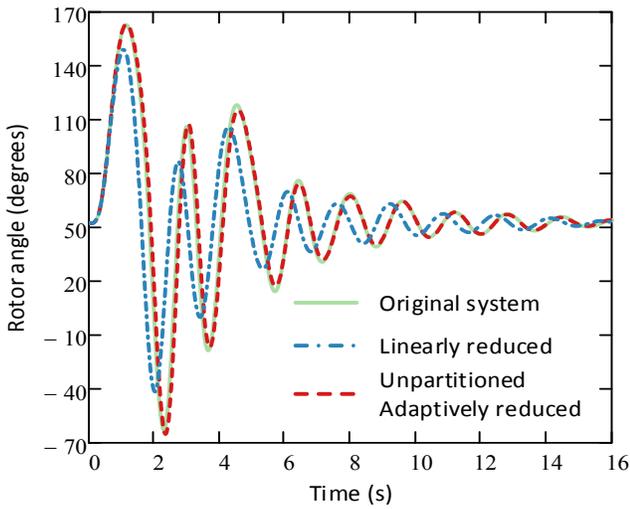

Fig. 5. Rotor angle of generator 23 following the fault at bus 3.

The fault at bus 3 is the largest disturbance of the NPCC system and generator 23 has the largest rotor angle error following this fault, and thus its rotor angle is used for the comparison. The results of simulation are shown in Fig. 4 when the adaptive approach is applied to partitioned system and in Fig. 5 when the adaptive approach is applied to unpartitioned system. As it can be seen from Fig. 4 and Fig. 5, if the external system is reduced using linear model reduction, the rotor angle trajectory differs significantly from the rotor angle trajectory of the original system whereas the rotor angle trajectory of the system reduced by the adaptive approaches follows accurately the original trajectory.

To present in more detail the difference between approaches, the root mean square errors of all states are calculated using the following expression:

$$\varepsilon_i = \sqrt{\frac{\sum_{j=1}^{N}(x_{ij} - \hat{x}_{ij})^2}{N}}, \quad (19)$$

where $N$ is the number of simulation steps; $x_{ij}$ and $\hat{x}_{ij}$ are respectively the values of $i$-th state variable of generator 23 of the original system and the reduced system at time step $j$.

The results of calculation are shown in Table II. The proposed adaptive approaches reduce the error by 74% to 81% for every state variable compared to the linear model reduction approach.

A fault with the duration equal to CCT is the worst-case scenario that can cause the largest error. The worst-case errors of the proposed adaptive model reduction approach are small enough to justify its applicability to power system stability studies.

TABLE II
COMPARISON OF ROOT MEAN SQUARE ERROR OF STATES OF GENERATOR 23

| States | System | | |
|---|---|---|---|
| | Linearly reduced | Partitioned adaptively reduced | Unpartitioned adaptively reduced |
| $\delta$, degrees | $2.24 \times 10^1$ | $5.5 \times 10^0$ | $5.1 \times 10^0$ |
| $P_m$, p.u. | $1.8 \times 10^{-3}$ | $4.1 \times 10^{-4}$ | $3.4 \times 10^{-4}$ |
| $P_{gv}$, p.u. | $2.3 \times 10^{-2}$ | $5.1 \times 10^{-3}$ | $4.5 \times 10^{-3}$ |
| $V_R$, p.u. | $1.7 \times 10^{-1}$ | $3.7 \times 10^{-2}$ | $4.1 \times 10^{-2}$ |
| $R_f$, p.u. | $1.3 \times 10^{-2}$ | $3.4 \times 10^{-3}$ | $3.2 \times 10^{-3}$ |
| $E_{fd}$, p.u | $9.8 \times 10^{-2}$ | $2.2 \times 10^{-2}$ | $2.3 \times 10^{-2}$ |
| $E'_d$, p.u. | $6.9 \times 10^{-2}$ | $1.4 \times 10^{-2}$ | $1.5 \times 10^{-2}$ |
| $E'_q$, p.u. | $1.1 \times 10^{-2}$ | $2.8 \times 10^{-3}$ | $2.6 \times 10^{-3}$ |
| $\omega$, p.u. | $4.7 \times 10^{-3}$ | $1.0 \times 10^{-3}$ | $9.7 \times 10^{-4}$ |

In addition to the accuracy the approaches are compared in terms of simulation time as shown in Table III. In Table III the large disturbance corresponds to the fault at bus 3 and the small disturbance corresponds to the fault at bus 17. The results show that the proposed adaptive approach applied to the partitioned system reduces the simulation time by 57% during the large disturbance and by 59% during the small disturbance compared to the original system. If the adaptive approach is applied to the unpartitioned system the simulation time is reduced respectively by 73% and 84%. Better speed performance of the second version of the adaptive approach is caused by the fact that the whole system is switched to the linearized model which is the fastest model. Especially it is clear in the case of the small disturbance as the switching to the linear model happens earlier.

Thus, the proposed adaptive approach provides both high accuracy and high simulation speed.

TABLE III
COMPARISON OF SIMULATION TIME

| System | Simulation time, seconds | |
|---|---|---|
| | Large disturbance | Small disturbance |
| Original unpartitioned | 3.7 | |
| Partitioned and linearly reduced | 0.9 | |
| Partitioned and adaptively reduced | 1.6 | 1.5 |
| Unpartitioned and adaptively reduced | 1.0 | 0.6 |

*B. Test of operating condition change*

To test how robustly the adaptive approach performs against a change of the operating condition, the temporary fault at bus 3 representing the largest disturbance is changed to a permanent fault cleared by tripping one of the lines connected to the bus or by complete isolation of the bus from the system by tripping all lines connected to the bus. Following the bus fault, the operating condition of the system is changed. The results of simulations are shown and compared in Table IV, from which the proposed adaptive approach can reduce the system after the change of the operating condition maintaining the accuracy of simulation.

TABLE IV
COMPARISON OF FAULTS LEADING TO A NEW OPERATING CONDITION

| Fault | Rotor angle RMS error, degrees | |
|---|---|---|
| | Partitioned adaptively reduced | Unpartitioned adaptively reduced |
| Temporary fault at bus 3 | 5.51 | 5.12 |
| Fault at line 3-2 followed by the line trip | 5.48 | 5.09 |
| Fault at line 3-4 followed by the line trip | 5.28 | 4.80 |
| Fault at bus 3 followed by the bus trip | 2.74 | 2.04 |

IV. CONCLUSIONS

This paper has proposed two versions of a new adaptive model reduction approach that can be applied to traditional partitioned system or to unpartitioned system for fast power system simulation. The approach is capable of accurate representation of the original power system model with significant reduction in computational time.